# Developing Tutorials for Advanced Physics Students: Processes and Lessons Learned

Charles Baily*[†], Michael Dubson[†] and Steven J. Pollock[†]

*School of Physics and Astronomy, University of St Andrews, St Andrews KY16 9SS Scotland, UK
[†]Department of Physics, University of Colorado, Boulder, CO 80309-0390 USA

**Abstract.** When education researchers describe newly developed curricular materials, they typically concentrate on the research base behind their design, and the efficacy of the final products, but do not highlight the initial stages of creating the actual materials. With the aim of providing useful information for faculty engaged in similar projects, we describe here our development of a set of in-class tutorials for advanced undergraduate electrodynamics students, and discuss factors that influenced their initial design and refinement. Among the obstacles to be overcome was the investigation of student difficulties within the short time frame of our project, and devising ways for students to engage in meaningful activities on advanced-level topics within a single 50-minute class period. We argue for a process that leverages faculty experience and classroom observations, and present several guidelines for tutorial development and implementation in upper-division physics classrooms.

## INTRODUCTION

A common theme in physics education research (PER) is that students will learn more when they are active participants in the classroom. [1] *Tutorials in Introductory Physics* [2] from the University of Washington (UW) is a prominent example of research-based activities that can measurably improve student learning. These tutorials were developed through an extended process of identifying and investigating common student difficulties, followed by several iterations of implementation, assessment and refinement. [3] Such large-scale projects typically involve student populations that number in the thousands, and can span multiple institutions.

Active-learning instructional strategies have also been used in physics courses beyond the introductory level, [4-6] and in-class activities have been a common feature of these curricula. However, when designing tutorials for advanced students, it may not be possible to exactly emulate the approach used at UW, due to institutional and funding constraints, as well as other factors particular to advanced physics courses. We discuss some of these obstacles, and our strategies for overcoming them during the process of transforming an upper-division electrodynamics course. [7, 8]

At the University of Colorado Boulder (CU), advanced undergraduate electromagnetism is a two-semester sequence: the first semester is devoted to electrostatics (E&M 1), and the second covers electrodynamics (E&M 2), where Griffiths is the usual textbook. [9] Both are offered every semester in the form of three 50-minute lectures per week, with a typical enrollment of 30-50 students. In the summer of 2011, following years of research and development work on E&M 1, [6] we secured funds to extend these efforts that were sufficient to support a post-doctoral researcher for two years, which set a basic timeline for the project. We arranged for ourselves to teach E&M 2 in the Fall 2011 (FA11) and Spring 2012 (SP12) terms, so there would be at least two sequential semesters in which to test and refine new materials.

## PRELIMINARY CONSIDERATIONS

Among our project goals was the creation of a set of classroom activities that would enhance student understanding of important topics in electrodynamics. While introductory-level tutorials are normally used in recitation sections that are separate from lectures, it was not possible to alter the structure of E&M 2 to include an additional recitation, because the number of discipline-specific credit hours required of physics majors is already at the maximum allowed by CU. The E&M 1 tutorials have previously been used in optional weekly sessions, but this was ruled out for our project, partly by ethical considerations – there have been demonstrable learning benefits for those who did participate in the tutorial sessions, [10] but not all students could attend due to scheduling conflicts. In fairness to all students, and in the interest of creating materials that could be used by other instructors facing similar dilemmas, the electrodynamics tutorials were designed for use during regular class periods.

The pre-existing research base on student difficulties in advanced electrodynamics was sparse (the only advanced E&M assessment known to us at that time [11] has no published results, and was more appropriate for engineering students). Without the possibility of conducting extensive investigations prior to the FA11 term, our best available resource for insights into common student difficulties was anecdotal reports gathered through individual one-hour interviews with six instructors who had recently taught E&M 2 at CU; this was supplemented by extended discussions with nine outside faculty members with experience in PER and curriculum development. [7]

Several themes emerged from these reports: Students had difficulty remembering the four Maxwell equations, in either differential or integral form (and with translating between the two representations). They had trouble physically interpreting mathematical expressions of conservation laws, such as the continuity equation. Many could not derive boundary conditions for the fields, and struggled with applying them in simple situations (e.g., no free charges or currents). Although they had used complex exponentials in other courses, their facility with them was weak, particularly in translating from complex quantities to ones that are physically measureable. They also had difficulty unpacking the numerous vectors appearing in plane waves. Reflection & Transmission is a topic where most of the issues listed above above can manifest simultaneously.

These reported student difficulties, along with others anticipated from having transformed E&M 1, were used to create a protocol for one-on-one interviews with students who had recently completed the E&M sequence. In contrast to our introductory courses, there were far fewer advanced students available for interviews, so we would be unable to draw definite conclusions about the prevalence of these issues in larger populations. It was also unclear whether the difficulties we observed months after the semester had ended could be attributed to not having learned the material well in the first place, or failing to retain that understanding over the long term. Although the five student interviews conducted that summer were useful for confirming most of the reported difficulties, it was apparent that the ideal research subjects would be students who were being exposed to topics in advanced electrodynamics for the first time.

An obvious drawback in continuing to explore student thinking during FA11, and waiting until SP12 to try out tutorials based on our observations, was the lack of direct student feedback prior to implementation regarding overall utility, and the clarity of the problem statements and diagrams. Another concern was time constraints: advanced physics problems generally require more time than introductory tasks, and we have seen that many students could not complete the tutorials developed at CU for electrostatics [12] and quantum mechanics [13] within 50 minutes.

Based on these considerations, we decided to create tutorials on a weekly basis during the FA11 semester, and immediately test them with a group of students from that class. These focus-group interviews would allow us to collect data from multiple students at the same time, while also observing how they worked together on these collaborative activities.

## FOCUS-GROUP INTERVIEWS

We recruited students from the FA11 class by offering $100 in exchange for 12 one-hour sessions, to be paid at the end of the term. There were only three volunteers, but these students happened to represent fairly well the range of abilities among their classmates: one was high performing; the second was average, but liked to throw out various ideas and think them through when tackling a problem; the third struggled throughout the term. The meetings took place every Friday morning, before the students had begun that week's homework assignment. Providing food for each session (costing $10/week) helped to keep their weekly attendance consistent.

When drafting the tutorials, we considered responses to concept tests, student-initiated questions, preflight submissions [14] and input from two undergraduate Learning Assistants [15] who interacted with students during class. Analyzing homework was not immediately useful in this situation because assignments were due long after we had moved on to other topics. Not every activity was based entirely on these observations – many articles in the *American Journal of Physics* proved to be useful sources of inspiration. [8, 16]

During these sessions, the three students were encouraged to think aloud and to talk with each other as they worked. The interviewer observed silently as much as possible, but asked clarifying questions when needed, and provided additional information when it was obvious the problem statements or diagrams were causing confusion. Otherwise, Socratic questioning was used when they felt stuck as a group, as would be normal in a tutorial setting. Some activities elicited confusion in unanticipated areas, causing us to change the focus of the tutorial to target these difficulties. Due to space limitations, we refer the reader to our online archive of course materials for examples of tutorial activities, and to the accompanying notes for instructors for specific student difficulties. [8]

These experiences led us to develop several general strategies for creating activities that are tractable for advanced physics students within a single class period:

**(i) Focus on concepts.** Students were easily slowed by complicated calculations. It proved just as useful to have them determine the sign of a certain quantity, or whether it was zero or non-zero.

**(ii) Scaffold the problems.** Tasks should progress in difficulty, with the initial problem statements being as explicit as possible, and successive problems less so, by building on information acquired in previous tasks.

**(iii) Don't underestimate completion times.** It was all too easy to convince ourselves that students would quickly finish tasks that appeared trivial to us. Students seemed to always require at least 10 minutes for every page of activities when adequate space had been provided for their written work.

## INITIAL IMPLEMENTATION

The FA11 interviews resulted in a set of 13 tutorials that were used in the SP12 semester; another nine short activities were added to cover review material and other basics, such as completing a derivation that would have otherwise been done by the instructor at the board. Approximately 40% of the lectures incorporated small-group work, which was facilitated by using a room with tables instead of individual desks. Depending on the specific activity, the completion times ranged from 10 to 45 minutes, and we typically oriented students to a tutorial with concept tests and some discussion. Implementing these activities presented challenges we didn't anticipate from our experience with introductory tutorials, and we describe here a few lessons learned in the process.

**(1) Sell students on group work.** Students have their own ideas about what an upper-division classroom should be like, and may at first be reluctant to engage in small-group activities, *particularly* if they view tutorials as only appropriate for introductory students. This mirrors a previous study of student attitudes prior to using concept tests in advanced physics courses. [17] Instructors can encourage peer discussion by reminding students that scientific argumentation is a skill that is developed through practice.

**(2) Pass out worksheets just prior to beginning.** When we gave them out at the start of class, there were inevitably some students who began working on the problems right away, even though we had not finished lecturing. This discouraged collaborative work later when students who were ahead of others at their table were reluctant to pause or start over again.

**(3) Create challenge problems.** With students working at their own pace, there were some groups that finished long before others. To keep these students using their time productively, we began to include one or two challenge questions at the end of each activity.

## RESULTS & DISCUSSION

Student reactions to the use of in-class tutorials in SP12 were generally positive, but there were some students who did not care at all for this style of learning. Both extremes are illustrated by comments from two students in an end-of-term survey:

*"I really like the tutorials! They gave me practice working with important ideas, and encouraged me to stay engaged."*

*"The tutorials were somewhat helpful with regard to my ability to retain the material, but I felt they were a colossal waste of class time. I didn't expect to spend my tuition money on hours of class time doing worksheets instead of directly learning from my professor."*

The latter comment underscores the importance of point (1) in the previous section – despite our best efforts, some students may still subscribe to a transmissionist theory of learning, and instructors should be aware that even those who liked using tutorials in their introductory courses might feel differently in an upper-division context.

In the same survey (given in SP12 and the following FA12 term), students were asked to rate various elements of the course on a scale of 1-5 for being *useful for their learning* (Fig. 1), and how much they *enjoyed using them* (Fig. 2), where "enjoyment" was meant as academic or intellectual pleasure.

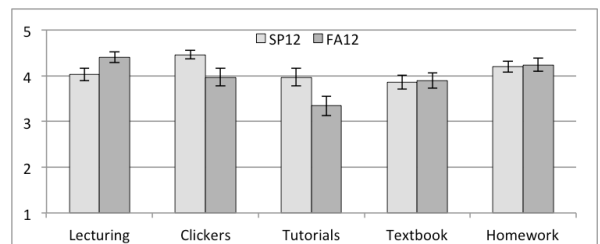

**Figure 1.** Post-survey results for SP12 & FA12 semesters. Students rated each course element in terms of **how useful it was for their learning** (5 = very useful; 1 = not at all useful). Error bars represent the standard error on the mean.

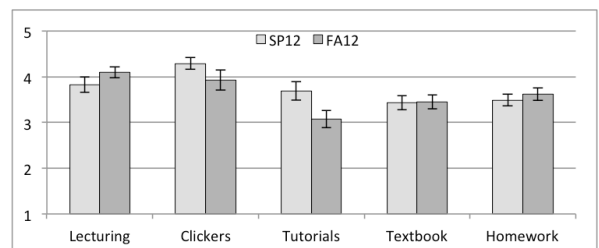

**Figure 2.** Post-survey results for SP12 & FA12 semesters. Students rated each course element in terms of **how much they enjoyed them** (5 = strongly like; 1 = strongly dislike). Error bars represent the standard error on the mean.

In SP12 (N=35), the perceived utility of tutorials (4.0±0.2) was comparable with other course elements, with the exception of the highly rated clicker questions (4.5±0.1). However, in the FA12 term (N=29), where only a subset of the tutorials and concept tests were used, both tutorials (3.3±0.2) and clickers (4.0±0.2) rated significantly lower on utility than pure lecturing (4.4±0.1). In every instance, the mean rating for "utility" was higher than for "enjoyment", though the difference was only statistically significant for the *textbook* and *homework* categories ($p < 0.05$ by a one-tailed t-test). There was a strong correlation between the two ratings in every category; for the tutorials specifically, $r = 0.90$ (SP12) and $r = 0.78$ (FA12).

It's possible some of the variance between semesters can be explained by differences in instructor practices, in that the non-PER instructor for FA12 used fewer concept tests and tutorials, and attended less to student affect. A preliminary implication is that the modes of instruction that were most prevalent in each course received the higher ratings, and that emphasizing the benefits of group work can positively impact student enjoyment and perceived utility.

There were also differences between SP12 and FA12 in terms of learning outcomes, as measured by the CURrENT [7, 8] electrodynamics assessment (Fig. 3). Average scores were roughly comparable on some questions, but moderately or significantly different on others. However, rigorous validation of the assessment hasn't yet been completed, and the number of students represented is fairly small. These data are therefore only tentative, and definite conclusions regarding the relationship between student learning and using tutorials requires further investigation.

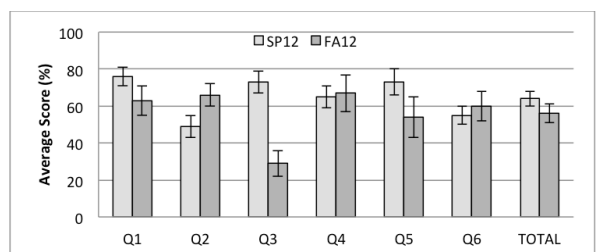

**Figure 3.** Results from the CURrENT conceptual assessment (Version 3) for SP12 (N=24) & FA12 (N=14) for individual questions and total score. Error bars represent the standard error on the mean.

These results suggest some parallels with findings from a previous study [18] at CU on student perceptions of introductory tutorials: even in classes with significant learning gains, students generally did not give high ratings on the utility of tutorials, and rated their enjoyment of them even lower. In that study, variations across semesters also implied that instructor practices were an important factor.

As for implementation styles, FA12 students expressed their preferences among three options. The average ratings were identical for occasional in-class tutorials and for optional weekly sessions; lower preference was given for having one class each week dedicated only to tutorials (a format that has been used in other advanced physics courses at CU).

The systematic process of course transformation requires an iterative approach that builds on a research base of student difficulties. Given the time constraints and relatively small class sizes in our upper-division courses, we have argued here for an initial process of tutorial development that leverages faculty experience, student feedback, classroom observations and ongoing interviews. This has led to heuristic guidelines for developers and implementers that should prove useful in the ongoing efforts at upper-division course transformations.

## ACKNOWLEDGMENTS

We gratefully acknowledge the many students and faculty members who have made our work possible, in particular: B. Ambrose, A. Becker and the PER@C group. Supported by the University of Colorado and NSF-CCLI grant #1023028.